\def\eqn#1{eq.~(\ref{#1})}

\documentclass[11pt]{article}
\usepackage{moriond,epsfig}

\def\be{\begin{equation}}
\def\ee{\end{equation}}
\def\bea{\begin{eqnarray}}
\def\eea{\end{eqnarray}}

\begin{document}
\vspace*{4cm}
\title{Theoretical Summary of Moriond 2004: \\ QCD and Hadronic Interactions}

\author{Andrzej Czarnecki}

\address{Department of Physics, University of Alberta\\
Edmonton, AB,   Canada T6G 2J1\\
and \\
TRIUMF, 4004 Wesbrook Mall, Vancouver, BC, Canada V6T 2A3}

\maketitle\abstracts{ Theoretical talks and discussions at {\em
Rencontres de Moriond 2004 ``QCD and Hadronic Interactions''} are
summarized.  Following exciting recent experimental discoveries,
theoretical developments were reported in the description of heavy
ion collisions, light hadron spectroscopy, and the physics of
hadrons containing heavy quarks.  Some of the predictions have
already been tested by subsequent experiments, notably by the very
recent SELEX observation of a possible new charm-strange meson
$D_{sJ}(2632)$.}

\section{Introduction}
The year between Recontres de Moriond in 2003 and in 2004 was rich
in exciting discoveries challenging some established views on QCD
and hadronic interactions: evidence of exotic behavior in heavy
ion collisions, sightings of pentaquarks, and new charmed
particles, just to name the most publicized ones. Some of these
phenomena had been predicted or speculated about by theoreticians,
and all fuelled interesting discussions at the QCD Moriond 2004.

In this summary, topics are organized according to increasing
``hardness", or decreasing characteristic distance scale of
phenomena.  Thus, we start with heavy ion collisions and QCD
evolution equations,  followed by properties and interactions of
light hadrons, and finish with a review of the recent progress in
charm spectroscopy and the physics of hadrons containing the $b$
quark.

\section{Heavy ion collisions: new phase of QCD?}
Experiments at the Relativistic Heavy Ion Collider (RHIC) probe
nuclear matter in a new regime of high temperature and density.
Heavy ion collisions have been studied extensively in the past,
most recently at the Super Proton Synchrotron (SPS) at CERN, where
experimenters announced circumstantial evidence for a new state of
matter in 2000.

One novel feature of RHIC is that the energy of individual parton
collisions is sufficiently higher than at the SPS for jets to be
produced and detected.  Observation of jets, and their dependence
on the colliding particles and the centrality of collisions,
probes the state of hadronic matter under extreme conditions.

Particularly striking evidence that something new occurs in the
nuclear medium after the collision is seen in the so-called
suppression of back-to-back jet correlations, discovered during
the last year and reviewed at this
meeting.\cite{d'Enterria:2004fm,Snellings:2004ep}  In the former
reference, in Fig.~4 we see a typical configuration of an
off-center collision of large nuclei.  The right hand side  of
that Figure shows angular correlations of dijet events.

In the case of proton-proton collision, conservation of momentum
forces the two jets to appear back-to-back.  This is represented
by the two peaks in that figure, separated by $180^\circ$.

In collisions of gold-gold nuclei, the overlap region is elongated
in the direction perpendicular to the plane of velocities of
incoming nuclei (reaction plane).  Thus, jets emitted within that
plane have a relatively short path to leave the overlap region.
Although they still have to cross a region of normal nuclear
matter, it does not  affect them significantly, and the
back-to-back correlation is preserved (a lower peak in the
$180^\circ$ correlation).

The intriguing phenomenon is observed for the ``out of plane"
jets.  When one jet is observed in the direction (roughly)
perpendicular to the reaction plane, none is observed in the
opposite direction! It seems to get totally absorbed, presumably
by interactions with the putative new state of QCD matter that
arises in the overlap region.

No similar suppression of the back-to-back jet correlations has
been observed in collisions of small projectiles with gold
(deuterium-gold).\cite{Adams:2003im}  This is interpreted as a
clear evidence that the suppression is a final-state effect, due
to interactions with the post-collision medium, rather than an
initial state effect, such as nuclear shadowing.

Explaining the mechanism of the jet attenuation is a challenge for
theorists.  Various approaches to describing hadron interactions
with the medium have been discussed and can be found in these
proceedings.\cite{Capella:2004zq,Moriond04Kharzeev,Moriond04Sousa}  Other probes
of the dense medium include gluon radiation off heavy quarks
\cite{Armesto:2004hz} and prompt photons.\cite{Arleo:2004ri}

\section{Evolution equations and resummation techniques}
Evolution equations are a source of information about physics
outside regions where fixed order perturbation theory is
appropriate.  In this meeting, their applications were seen in
several contexts, from the small $x$ limit, to heavy flavor
production (see Section \ref{beauty}), to Higgs production in the
highest energy collisions.

In the limit of very high center-of-mass energy, $\sqrt{s}$, the
BFKL equation describes effects of large logarithms in the ratio
of the (fixed) transferred momentum to $\sqrt{s}$.  This regime is
of significant practical interest, for example because gluon
exchanges give rise to multi-jet events through the BFKL dynamics,
and create a background for the Large Hadron Collider (LHC)
searches for new heavy particles, and in studies of known
particles such as the top quark.

In the leading order (LO),  various techniques are available for
solving this equation, thanks to the conformal symmetry of the
BFKL kernel.  This symmetry is spoiled in the next-to-leading
(NLO) order, because of the running coupling constant, and the
solution becomes a challenge.  Recently, a solution method at the
NLO has been
found.\cite{Andersen:2004by,Andersen:2003an,Andersen:2003wy} It
expresses the solution as a phase space integral, thus controlling
the energy-momentum conservation.  This is an important
breakthrough, clarifying previously found paradoxes in the
behavior of the NLO solutions, and opening a way to
phenomenological studies of the BFKL dynamics.

As the cross section grows with increasing energy, the BFKL
evolution must be supplemented with non-linear effects.  The
number of partons cannot grow without limit, and a so-called
saturation is reached.  Near the saturation regime, another
equation has been proposed by
Kovchegov.\cite{Kovchegov:1999ua,Kovchegov:1999yj}  At this
meeting, recent improvements on this formalism, including effects
of fluctuations, were reported.\cite{Mueller:2004qj}

The high energy collisions at the LHC will likely result in
production of the Higgs boson.  Two mechanisms of the Higgs
production were discussed.\cite{Moriond04Demine,Grazzini:2004qx}
In the leading production channel, through a gluon fusion, it is
important to determine the perturbative QCD corrections to the
transverse momentum ($q_T$) distribution of  Higgs bosons.  In the
majority of events, Higgs will have a relatively small transverse
momentum, $q_T^2\ll M_H^2$, and the distribution obtained in a
fixed order perturbative calculation will be distorted by large
logarithms $\ln(M_H^2/q_T^2)$.  The state of the art in resumming
such effects has been presented.\cite{Grazzini:2004qx}  The QCD
description includes now next-to-leading order corrections to the
distributions, improved by a resummation of
next-to-next-to-leading logarithms.\cite{Bozzi:2003jy}

\section{Light hadrons: real and virtual}
\subsection{Pentaquarks}
Exotic baryons, with a flavor structure more complex than can be
built with three quarks, were searched for by a number of
experiments, especially in the 1970's, always with negative
results. Since no evidence of their existence was found,
theoretical models predicting their existence were treated with
suspicion.  Eventually, the experimental searches were given up,
until recently.

Since the last year, we have been witnessing a revolution. First,
the pentaquark $\Theta^+(1540)$ was discovered, after a prediction
of its mass \cite{Biedenharn:1984su,Praszalowicz:1987em,Diakonov:1997mm} and a
narrow width:\cite{Diakonov:1997mm} a remarkable success for both
theory and experiment.  Further, cascade pentaquark states
$\Xi^{--}$ and $\Xi^{0}$ were reported. Four such particles are
expected to form an isospin quartet.  The doubly-negative and the
neutral ones can decay into final states with all particles
charged and are therefore somewhat easier to identify, but even
for them the evidence is scarce.   In addition, the antiparticle
of $\Theta^+$ has been detected by one experiment. Experimental
situation is not yet entirely clear: some experiments do not see
states that others do, and mass determinations vary. This is
summarized in Marek Karliner's
contribution.\cite{Moriond04Karliner}  Lack of confirmation by
some experiments may be interpreted as evidence of particular
production mechanisms.\cite{Karliner:2004gr}

The nature of the pentaquark states has been vigorously disputed
since their reported discovery. It does not seem possible to
explain the low mass and narrow width of $\Theta^+$ in the
constituent quark model.\cite{Close:2003tv}  The mass prediction
for a $uudd\bar s$ state (the lowest quark flavor content
consistent with $\Theta^+(1540)$ decays) is about 1800 MeV.  Its
width would be expected in the ballpark of 100 MeV, while the
experiments suggest one or two orders of magnitude smaller width.

Among the more successful models, the main dividing line is
between the chiral soliton  and the correlated quark approaches.
Other interpretations have also been proposed and some were
discussed at this meeting.\cite{Bicudo:2004yp}  The chiral soliton
model addresses the non-trivial structure of the QCD vacuum.  Its
clear achievement is the prediction of the $\Theta^+$ that led to
its discovery.  The correlated quark model
\cite{Karliner:2003sy,Jaffe:2003sg} operates with quasiparticles,
constituent quarks.  Unlike in the traditional quark model, in
this approach quarks are  spatially correlated and form diquarks.
It has been speculated that the existence of diquarks may help
explain hard-core repulsion between nucleons and the $\Delta
I=1/2$ rule.\cite{Jaffe:2004zg}

Although the two approaches seem to be fundamentally different,
they were shown to be equivalent and connected to QCD in the limit
of a large number of colors, $N_c\to
\infty$.\cite{Jenkins:2004tm,Manohar:1984ys}  In fact, it has been
argued that the accuracy of either model does not extend beyond
where their predictions are the same.\cite{Jenkins:2004tm}

There is a number of outstanding questions regarding pentaquarks,
among which solidifying evidence of their existence is the most
important.  The production mechanisms, leading to surprisingly
large production cross sections, are a very interesting issue.
Some information about production mechanisms and wave functions
may come from searches in heavy meson decays.\cite{Browder:2004mp}
Determination of parity is often mentioned as very important; the
chiral soliton and correlated quark models all predict positive
parity.  The fact that several well-established baryons, including
some ground state cascades, have not had their parities measured
yet, shows that this may be a very difficult task.

There is also a tantalizing signal of a charmed pentaquark
$\Theta_c(3099)$, significantly heavier than predictions based on
analogies with $\Theta^+(1560)$, on the order of 2700 MeV.
Unfortunately, evidence for this state comes from only one
experiment (H1).  If it is confirmed, there may be a lighter
charmed pentaquark of which $\Theta_c(3099)$ is a so-called chiral
doubler of opposite parity.\cite{Nowak:2004jg} We will return to
this topic in Section \ref{sec:doublers}.

\subsection{Muon $g-2$}
In January 2004, the Brookhaven $g-2$ Collaboration announced a
measurement of the negative muon anomalous magnetic
moment,\cite{Bennett:2004pv}
\be a_{\mu^-}\equiv {g_{\mu^-}-2
\over 2}  = 116 \, 592 \, 140 (85)\times 10^{-11}.
\ee
This is
higher than the 2000 measurement with positive muons
\cite{Bennett:2002jb} by $100\times 10^{-11}$. This difference is
statistically insignificant and  the two values can be combined to
give the world average for the muon,
\be a_\mu = 116 \,592 \,
080 (60)\times 10^{-11}. \ee The upward shift of this average, due
to the latest $\mu^-$ data, increases the discrepancy between the
experiment and the Standard Model prediction.  At this meeting,
Arkady Vainshtein reviewed recent improvements in the most challenging
aspect of the latter, namely the hadronic effects.
\begin{figure}[htb]
\hspace*{20mm}
\begin{tabular}{c@{\hspace*{15mm}}c@{\hspace*{15mm}}c}
\epsfxsize=30mm \epsfbox{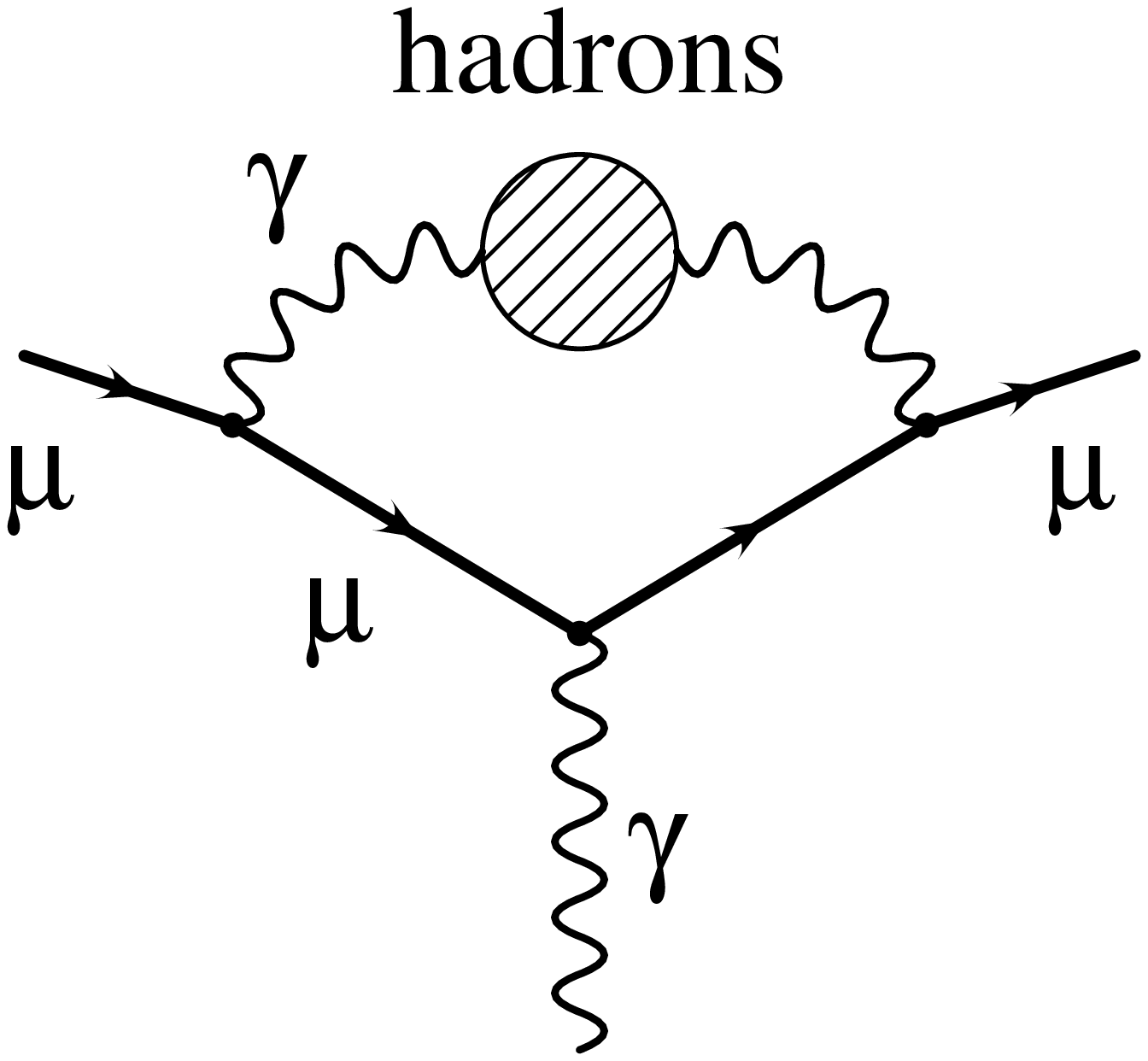} &
\epsfxsize=30mm \epsfbox{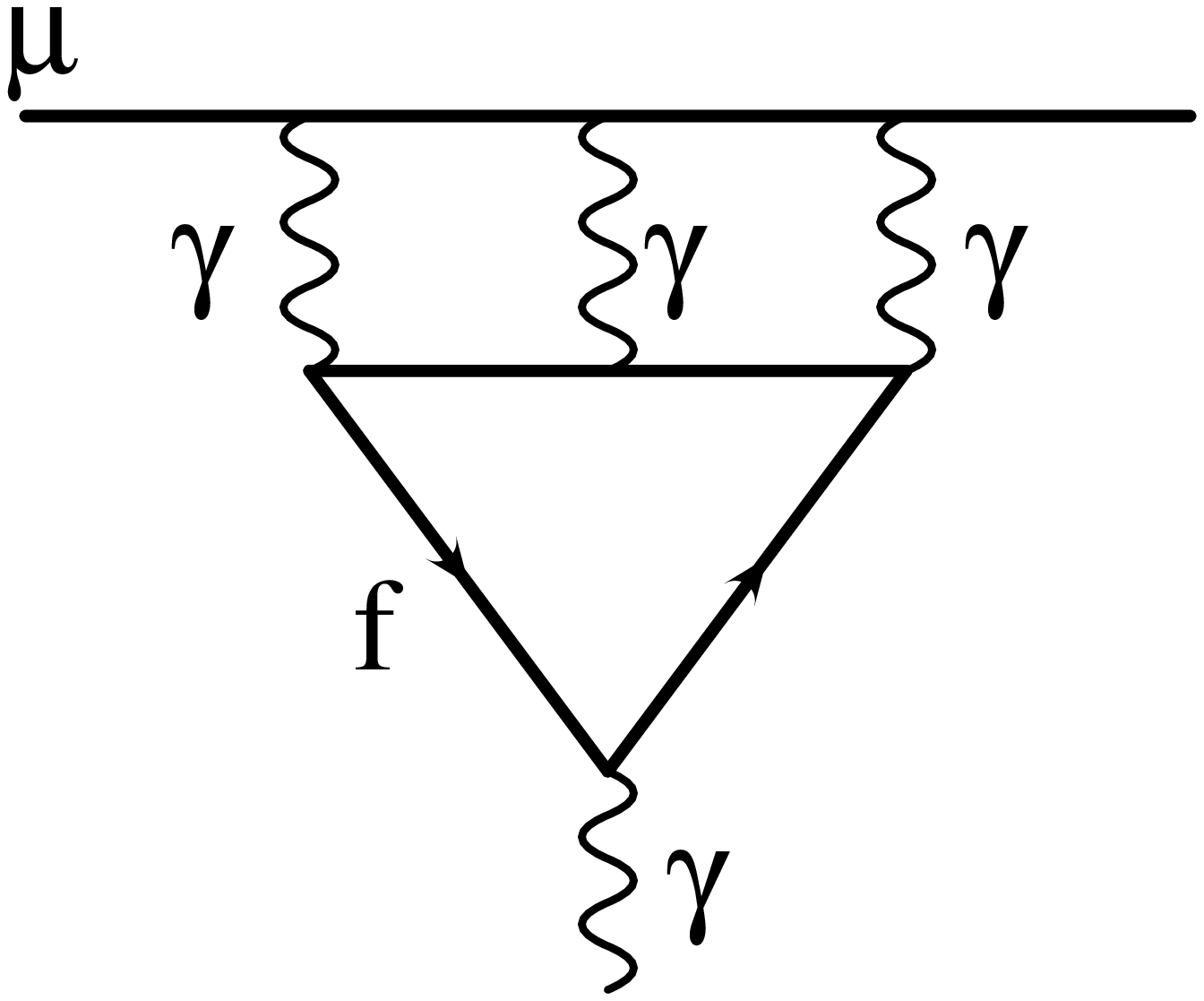} &
\epsfxsize=30mm \epsfbox{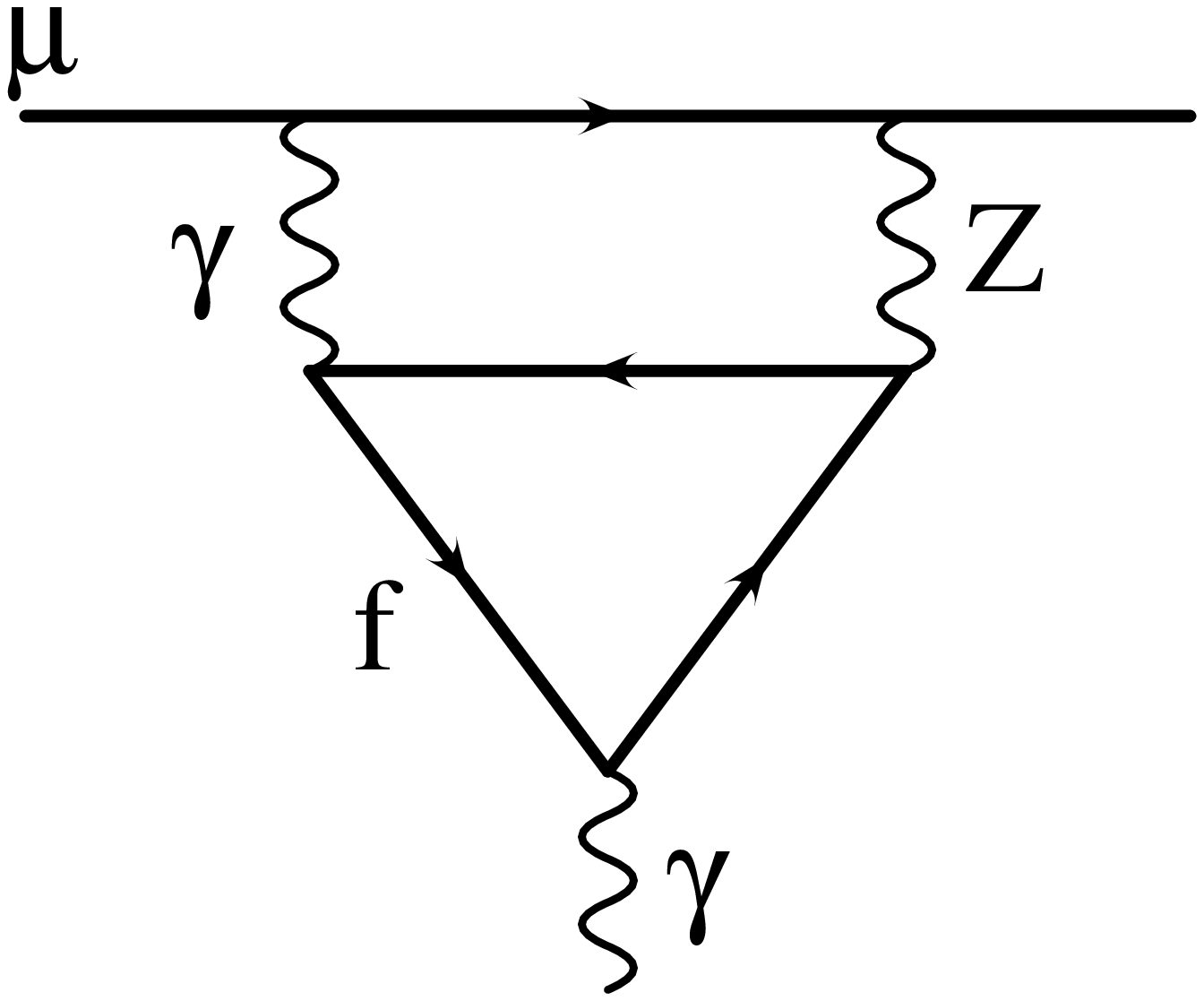}
\\
(a) & (b) & (c)
\end{tabular}
\caption{Examples of hadronic contributions to the muon anomalous magnetic moment:
(a) vacuum polarization, (b) light-by-light scattering, and (c) hadronic-electroweak contribution.
"f" denotes the quark species in the triangle loop.}
\label{fig:g2}
\end{figure}
All types of hadronic contributions to $g-2$, shown in Fig.~\ref{fig:g2},
have been subjects of interesting theoretical disputes and controversies in the last few years.
Among them, the light-by-light (LBL)
scattering part in Fig.~\ref{fig:g2}(b)
may set the ultimate limit for the accuracy of the Standard Model prediction.
Thus, its recent study by Melnikov and Vainshtein \cite{Melnikov:2003xd}
is an important new development.

Previous studies of this contribution were based on assumptions
about photon interactions with virtual pions (and other light
mesons), extrapolated from known interactions with on-shell pions.
That reasoning turns out to be not entirely correct.  The improved
analysis \cite{Melnikov:2003xd} employs the operator product
expansion and matching with exactly calculable asymptotic behavior
of QCD amplitudes. One finds that contributions of large
virtualities of loop momenta are less strongly suppressed than had
been previously assumed.  As result, the LBL contribution turns
out to be larger by about half than its previous estimates,
bringing the Standard Model prediction somewhat closer to the
Brookhaven experimental result. The theoretical prediction is
smaller than the experiment by about 2.2--3$\sigma$, depending on
the treatment of the vacuum polarization contribution in
Fig.~\ref{fig:g2}(a), obtained from $e^+e^-$ annihilation into
hadrons.

The vacuum polarization contribution is responsible for the major
part of the present theoretical uncertainty.  So far, it has been
estimated using $e^+e^-\to $ hadrons or hadronic decays of $\tau$
leptons.  The latter has excellent statistical accuracy in the low
energy region, crucial for the $g-2$, but suffers from systematic
theoretical uncertainties in the translation into $g-2$. The
Standard Model prediction based on the $\tau$ data tends to be
higher by about $150\times 10^{-11}$ than that based on $e^+e^-$,
and agrees much better with the measurement.  It is important to
understand the reason of the difference between $e^+e^-$ and
$\tau$ data, keeping in mind that the analysis of $e^+e^-$ is also
not entirely free from systematic uncertainties.  It now appears
\cite{Bill} that not all isospin breaking corrections have been
applied to the $\tau$ data.

At this meeting, Achim Denig presented new experimental results from
KLOE, that determine the low energy $e^+e^-\to $ hadrons
annihilation cross-section using radiative return.  Those data
support direct $e^+e^-$ annihilation measurements, and strengthen
the case for a revision of the analysis based on $\tau$ decays.

\subsection{The NuTeV puzzle}
Another recent experimental result disagreeing with the Standard Model is the measurement
of the weak mixing angle, $\sin^2\theta_W$, in muon neutrino and anti-neutrino scattering
on an iron target,  by the NuTeV Collaboration.\cite{Zeller:2001hh}
 We have heard two
talks on the QCD aspects of this discrepancy, from  experimental \cite{Mason:2004yf}
and theoretical perspectives.\cite{Kretzer:2004tv} The main focus of those studies is the possible
asymmetry in the energy carried by virtual strange quarks and anti-quarks in the nucleons.

\begin{figure}[htb]
\hspace*{28mm}
\begin{tabular}{c@{\hspace*{15mm}}c}
\epsfxsize=40mm \epsfbox{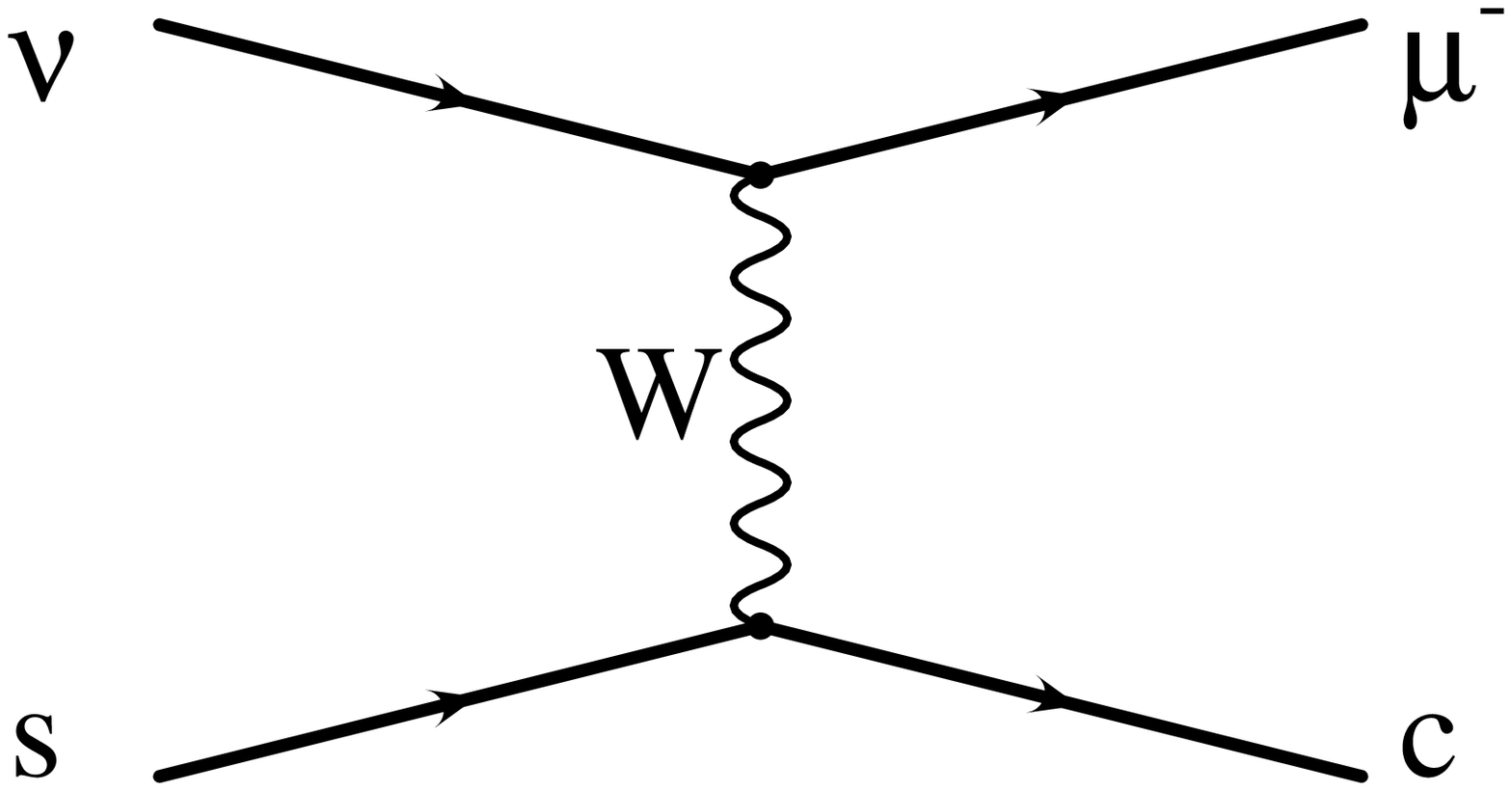} &
\epsfxsize=40mm \epsfbox{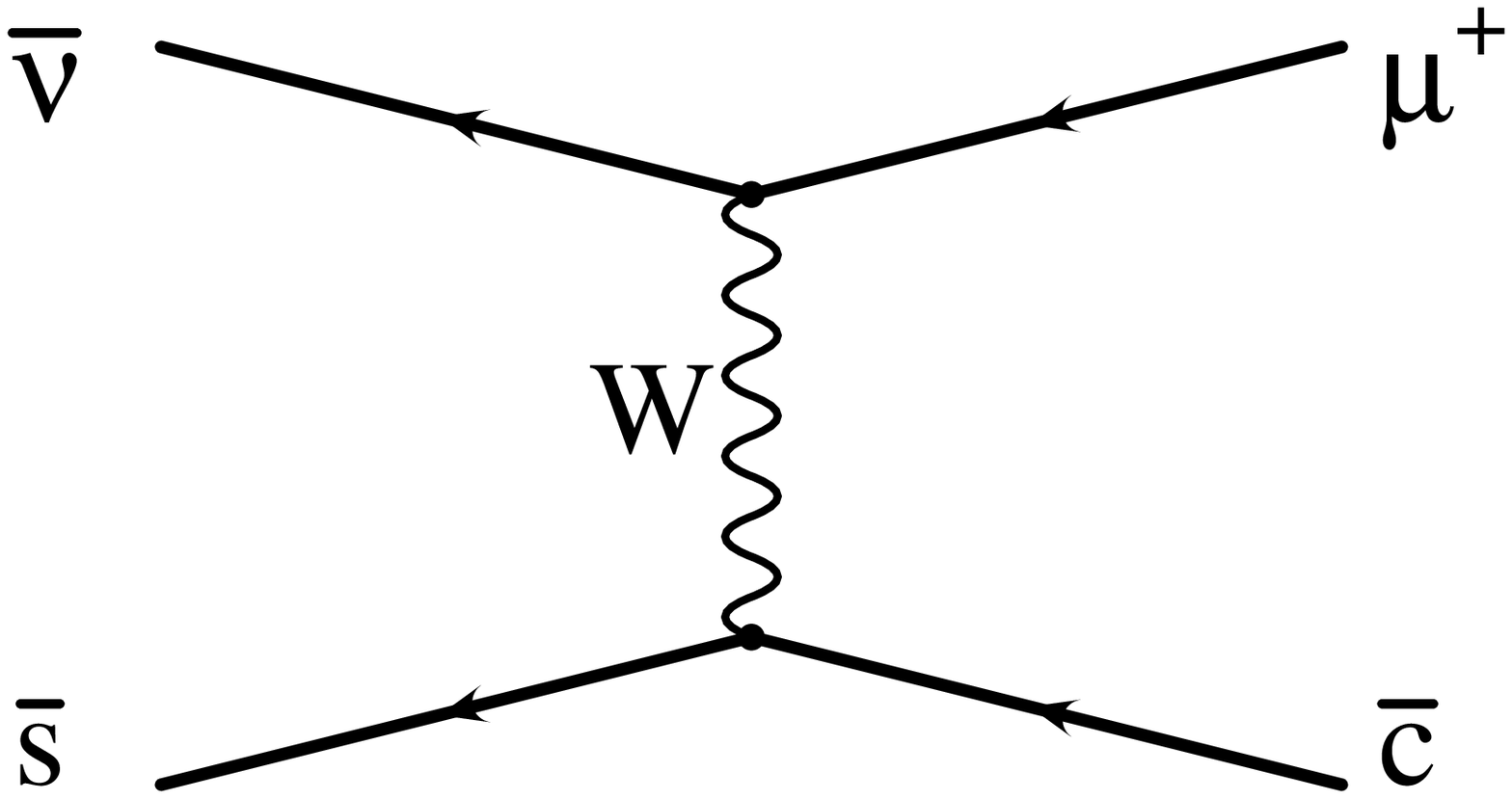}
\end{tabular}
\caption{Higher average energy of strange sea quarks compared to anti-quarks
enhances the charged current interactions
of neutrinos, relative to those of anti-neutrinos.}
\label{fig:NuTeV}
\end{figure}

It is easy to understand  how such asymmetry may influence the
$\sin^2\theta_W$ determination.  Consider the Paschos-Wolfenstein
ratio \cite{Paschos:1972kj} (it is related, but not identical, to
the cross-section ratios measured by NuTeV), \be R^- \equiv {
\sigma\left( \nu_\mu N \to \nu_\mu X\right) - \sigma\left(
\bar\nu_\mu N \to \bar \nu_\mu X\right) \over \sigma\left( \nu_\mu
N \to \mu^- X\right) - \sigma\left( \bar\nu_\mu N \to \mu^+
X\right) }= {1\over 2} - \sin^2\theta_W. \label{R} \ee Suppose now
that, for some reason, strange quarks carry, on average, a larger
fraction of the nucleon momentum than do strange antiquarks (see
Fig.~\ref{fig:NuTeV}).  Since the (anti)neutrino scattering
cross-section grows with the center of mass energy, such asymmetry
increases the denominator in \eqn{R} and decreases $R^-$.  This
decrease may be wrongly attributed to a too large value of
$\sin^2\theta_W$, if the strange sea asymmetry is present but
neglected.

Unfortunately, it is not yet clear whether this asymmetry is
present, and whether its magnitude and sign can help explain the
NuTeV puzzle.  An analysis of dimuon production, carried out by
the NuTeV collaboration, suggests that the asymmetry is very small
(consistent with zero), and with a tendency towards negative
values, that would even slightly increase the discrepancy between
the extracted value of $\sin^2\theta_W$ and the Standard Model
prediction.  On the other hand, a study by the CTEQ collaboration
found a positive asymmetry that can reduce the discrepancy.  It is
fortunate that CTEQ and NuTeV collaborate to clarify this issue.

\subsection{Other topics}
Other important topics concerning light hadrons were also
discussed.  Description of high energy forward hadronic scattering
was reviewed \cite{Pelaez:2004yv} and a discrepancy with previous
studies was pointed out.  Multiplicity distributions in
proton-(anti)proton and electron-positron collisions were
described. \cite{Dremin:2004wp} A very interesting analogy was
suggested between moments of those distributions and virial
coefficients in statistical mechanics, whose behavior is related
to phase transitions. Those results were illustrated with examples
using the Dual Parton Model and the Quark Gluon String Model, also
discussed in the context of production and decays of charmed
baryons.\cite{Piskounova} Also, investigations of the QED Compton
process in electron-proton collisions were
reported.\cite{Mukherjee:2004bh} All these studies presented
predictions testable at future experiments, LHC and eRHIC.

\section{Renaissance of the charm spectroscopy}
\label{sec:doublers}
An exciting series of discoveries of charmed hadrons has begun
shortly after the previous Moriond meeting.  Here I will focus on
the narrow $D_{sJ}$ states, of which the first $D_{sJ}(2317)$ was
reported by BaBar in April 2003, followed by confirmations by
other groups, and a discovery of $D_{sJ}(2460)$ by CLEO.
A number of interpretations have
been put forward to explain the narrow widths of those particles,
including baryonia, tetraquarks, etc.

At this meeting we heard a talk by Maciej Nowak, who with Rho and
Zahed \cite{Nowak:1992um,Nowak:1993vc} had predicted the existence
of so-called chiral doublers of heavy quark hadrons about ten
years prior to those discoveries.  Such prediction was also made
shortly afterwards by Bardeen and Hill.\cite{Bardeen:1993ae}  It
is very tempting to interpret the new states as a confirmation of
that prediction.  The chiral doubler picture is especially
attractive in that it predicts masses of further states and can be
tested by ongoing experiments.

To understand the idea underlying chiral doublers, consider
symmetries of the hadronic spectrum of states containing a heavy
quark.  One such symmetry is due to suppression of the
chromomagnetic interactions of the heavy quark and is known as the
Isgur-Wise or spin-flavor symmetry.\cite{Isgur:1991wq} One of its
manifestations is the small splitting between masses of mesons
such as $D$ and $D^*$, $B$ and $B^*$, or, of particular present
interest, $D_s$ and $D_s^*$; that is, between the ground state
$0^-$ and its hyperfine excitation $1^-$. This splitting vanishes
as the mass of the heavy quark tends to infinity.

The spectrum feature predicted in the chiral doubler picture is
due to the chiral symmetry. Each hadron containing a heavy quark
is supposed to have a partner (doubler) differing only by parity.
Now, the mass splitting does not vanish in the large mass limit,
because the chiral symmetry is broken (the splitting, like
constituent quark masses, is primarily due to the spontaneous
breakdown of the chiral symmetry). In the infinitely heavy quark
limit, the splitting between say a $D_s$ meson and its doubler is
the same as that between $D_s^*$ and its doubler. Thus, the chiral
doubler picture is very predictive in the $B_s$ system, yet to be
fully explored experimentally.  (Heavy mesons containing light
quarks $u,d$ instead of $s$ also have chiral doublers, but they
are generally broader, since they can decay by emitting a pion.)

\begin{figure}[htb]

\vspace*{4mm}
\hspace*{40mm}
\epsfxsize=80mm \epsfbox{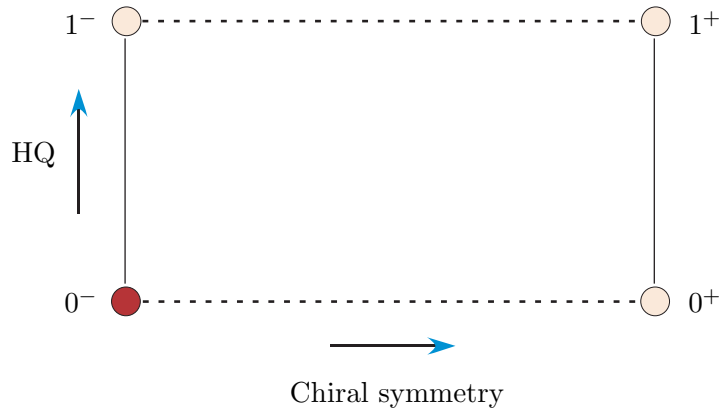}
\begin{picture}(0,0)(34,0)
\put(-55,26){HQ}
\put(-18,-6){Chiral symmetry}
\put(-48,6){$0^-$}
\put(-48,43){$1^-$}
\put(35,6){$0^+$}
\put(35,43){$1^+$}
\end{picture}
\vspace*{10mm}
\caption{Two symmetries for hadrons containing a heavy quark and light quarks:  Isgur-Wise
 or heavy quark (HQ) spin-flavor symmetry (vertical) and chiral symmetry (horizontal). Labels denote
 spin and parity of the states, with the $0^-$ being the ground state (dark circle).}
\label{fig:doubl}
\end{figure}

Within the chiral doubler picture, the quantum numbers of the lowest-lying $D_s$ mesons are
assigned as shown in Fig.~\ref{fig:doubl}.  The established states $D_s(1968)$ and $D_s^*(2112)$
correspond to the left-hand side of that picture, respectively to $0^-$ and $1^-$, and are
connected by the heavy quark symmetry.  The recent BaBar and CLEO states are interpreted as their
chiral doublers, $\widetilde D_s(2317)$ and  $\widetilde D_s^*(2460)$, and placed on that figure
as $0^+$ and $1^+$. The tilde denotes chiral partners.\cite{Nowak:1992um}
Note that the horizontal sides of this quadrangle are identical within error
bars,\cite{PDBook2004}
\bea
m_{0^+}- m_{0^-}=m_{\widetilde D_s}- m_{D_s} &=& 349.1(1.0) \mbox{ MeV},
\nonumber \\
m_{1^+}- m_{1^-} =m_{\widetilde D_s^*}- m_{D_s^*}&=& 347.2(1.5) \mbox{ MeV}.
\label{split}
\eea
The magnitude of these splittings is on the order of the chiral symmetry breaking parameters.
Their equality was predicted in the chiral doubler picture.\cite{Nowak:1993vc} The numerical value
is very close to the  prediction of 338 MeV.\cite{Bardeen:1993ae}

It is now very interesting to consider angular excitations of the light quark.  At the time of
this Moriond QCD, there were two
well established states $D_{s1}(2536)$ ($1^+$) and $D_{s2}(2573)$ (consistent with $2^+$).
The existence of their chiral partners was predicted,\cite{Nowak:2003ra,Moriond04Nowak}
within a smaller
mass difference than the ground state splittings in \eqn{split},
\bea
m_{\widetilde D_{s1}}= 2721(10)\mbox{ MeV}, \qquad
m_{\widetilde D_{s2}} =2758(10)\mbox{ MeV},  \qquad
 \mbox{(2003 prediction\protect\cite{Nowak:2003ra})}
 \label{pred}
\eea
Very recently (three months after the 2004 Moriond QCD), another charm-strange meson
$D_{sJ}^+(2632)$ has been discovered by the charm hadroproduction experiment
SELEX at Fermilab.\cite{Evdokimov:2004iy}  It has already been subject of several theoretical
interpretations, and may provide the ultimate test of the chiral doubler picture, particularly
that it is now possible to make a very precise prediction for another state.

\begin{figure}[htb]
\hspace*{40mm}
\epsfxsize=80mm \epsfbox{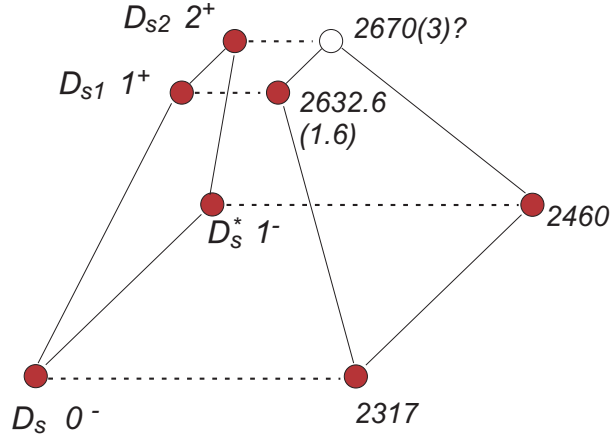}
\caption{The chiral doubler picture of charm-strange mesons as a Mayan pyramid.  The base consists
of states connected to the ground state by heavy quark and chiral symmetries, as shown
in Fig.~\protect\ref{fig:doubl}.  The upper level corresponds to the angular excitations of the
light quark.  The left hand side of that level is well established.  The state $D_{sJ}(2632)$ (also
called here $\widetilde D_{s1}$) has
been discovered most recently.  The open circle is the predicted state $\widetilde D_{s2}$.}
\label{fig:Mayan}
\end{figure}

If the SELEX discovery is confirmed,  it would fit the chiral doubler scenario as
$\widetilde D_{s1}$ (with the chiral splitting between $D_{s1}$
and $\widetilde D_{s1}$ even smaller than the prediction,
\eqn{pred}).\cite{Nowak:2003ra,Moriond04Nowak}  If we assume that the finite heavy quark mass
corrections are similarly small as in the ground state sector, see
\eqn{split}, there should exist another state, with the mass
\bea m_{\widetilde D_{s2}} =2670(3)\mbox{ MeV}.
\eea
The emerging picture of the
charm-strange mesons would resemble a Mayan pyramid, as shown in
Fig.~\ref{fig:Mayan}.  The horizontal distances between linked
states are supposed to represent their mass differences.

It is interesting to contrast the chiral doubler prediction with
other interpretations of the latest discovery $D_{sJ}(2632)$.  One
such interpretation is that it is an S-wave diquark-antidiquark
scalar.\cite{Maiani:2004xg,Nicolescu:2004in}  Explicitly, it would
be $[cd][\bar d \bar s]$ state, that would likely be accompanied
by a nearby charge +2 state,\cite{Maiani:2004xg} $[cu][\bar d \bar
s]$.  This is a clear, testable difference with the chiral doubler
picture, where the prediction is for another state of equal
charge, +1.

Another interpretation suggests that $D_{sJ}(2632)$ is a radial
excitation of the $D_s^*(2112)$.\cite{Chao:2004nb,Barnes:2004ay}
That would be an $S$-wave $1^-$ state.  If $D_{sJ}(2632)$ is a
chiral doubler of $D_{s1}$, the light quark is predominantly in a
$D$ wave (it is a pure $D$ wave in the limit $m_c\to \infty$).

The outstanding challenges for the chiral doubler approach, if
confirmed by a discovery of $\widetilde D_{s2}$, is to explain why
the chiral shifts of the excited mesons are so small, and to
explain the unusual decay pattern of the $\widetilde D_{s1}$ found
by SELEX.\cite{Evdokimov:2004iy}


\section{Production and decay of beauty}
\label{beauty}
 Over several years,  theoretical predictions
disagreed with measurements of $b$ quark production in hadronic
collisions. Next-to-leading (NLO) QCD calculations seemed to
significantly underestimate the inclusive cross section, by as
much as a factor of three.  This was puzzling, since inclusive
observables at high energy scales are expected to be reliably
described by perturbative QCD.  Recently, this problem has been
solved by a confluence of theoretical and experimental
developments.\cite{Cacciari:2003uh,Moriond04Cacciari}

Part of the solution was the availability of ``unprocessed" data,
from measurements by the CDF collaboration.  Those results were
expressed in terms of real hadronic final states ($B$ mesons),
avoiding as much as possible extrapolations and deconvolution,
that might have introduced biases from simulations.  This allowed
the theorists to employ $b$ quark fragmentation functions in a
fully controlled manner, in connection with an NLO QCD calculation,
including mass dependence, and supplemented by a resummation of
logarithms of the transverse momentum to mass ratio. The
fragmentation functions (a non-perturbative input) were obtained
from $e^+e^-$ annihilation data for the $b$ production. Another part
of the solution was that the disagreement between theory and
experiment turned out to have been somewhat exaggerated, because
theoretical uncertainties were not fully appreciated or accounted
for in comparisons.

A lesson  \cite{Moriond04Cacciari} drawn from this long story
of perceived disagreements is
that it is helpful if experiments  publish physical
observables, perhaps in addition to deconvoluted numbers expressed
in terms of quarks.  Such approach, giving
theorists unambiguous quantities to make comparisons with, may
help clarify remaining discrepancies in heavy quark production in
electron-proton and photon-photon collisions.

Theoretical progress was also reported in hadronic
\cite{Keum:2003js,Moriond04Lu} and semileptonic
\cite{Uraltsev:2004ta} decays of $B$ mesons.  In hadronic decays
$B\to DM$ and $B\to D^* M$, with $M$ denoting  a light meson, a
perturbative QCD formalism was employed to determine factorizable
and non-factorizable amplitudes, including strong phases.
Predictions are now available for $M=\rho,\omega$ and can be
tested by  measurements at $B$
factories.\cite{Keum:2003js,Moriond04Lu}

In semileptonic decays, a very impressive recent achievement is
the extraction of several heavy quark/meson parameters from
moments of lepton energy and hadronic mass.  Both the theoretical
and experimental efforts that were essential to this end were
reported at this meeting.\cite{Uraltsev:2004ta,Flaecher:2004sw} On
the theory side, a new analysis removed the previous disagreement
between predicted and measured dependence of the hadron mass
moments on the lepton energy cut.

\section{Outlook}

Thanks to the numerous recent experimental results, QCD and
hadronic interactions are again a fascinating research area. If
there is a common theme to the news from the diverse fields, from
heavy ion collisions, to light baryon spectroscopy, to heavy quark
hadrons, I believe it is the insight into the non-trivial
structure of the QCD vacuum.  Are we seeing the effects of the
spontaneous chiral symmetry breaking in the new narrow
charm-strange mesons?  Do exotic light states, predicted by
the chiral soliton model, really exist?  Is the deconfining phase transition
being seen in heavy ion collisions?  These are exciting questions
that may soon be definitively answered.  There is a flow of new
experimental results which challenge theorists' insight and
creativity.  There are also theoretical predictions that will be
confronted by forthcoming measurements.  The next Moriond meeting
promises to be very interesting, just like this one.

\section*{Acknowledgments}
For generous help in preparing this talk, I thank Jeppe Andersen,
Alfonse Capella, Marek Karliner, Dmitri Kharzeev, Yuri Kovchegov,
Stefan Kretzer, William Marciano, Kirill Melnikov, Maciej Nowak,
Bolek Pietrzyk, Robert Pisarski, Micha{\l} Prasza{\l}owicz,
Nikolai Uraltsev, Arkady Vainshtein, and Mikhail Voloshin.  I am
grateful to Stephen Godfrey, Marek Karliner, William Marciano, and
Maciej Nowak for reading this manuscript and suggesting
corrections.

Special thanks are due to the organizers of this conference, particularly Etienne Aug\'e
and  Jean Tran Thanh Van, for the excellent meeting.

This work was supported by the Science and Engineering Research
Canada (NSERC).

\section*{References}

\end{document}